\input harvmac

\def \ua {\uparrow}
\def \Q {{\cal Q}}

\def \k {\kappa}

\def \del {\partial}
\def \bd {\bar \partial }

\def \const {{\rm const}}
\def \ha{{\textstyle{1\over 2}}}

\def \chi {\chi}

\def \p {\phi}
\def \m {\mu}
\def \n {\nu}

\def \l {\lambda}

\def \td {\tilde }

\def \inv {^{-1}}
\def \ov {\over }

\def \lr { \lref}
\def\np {{  Nucl. Phys. }}
\def \pl {{  Phys. Lett. }}
\def \mpl {{ Mod. Phys. Lett. }}
\def \prl {{  Phys. Rev. Lett. }}
\def \pr  {{ Phys. Rev. }}

\def \cqg {{ Class. Quant. Grav. }}

\baselineskip8pt
\Title{
\vbox
{\baselineskip 6pt{\hbox{PUPT-1639}}{\hbox
{Imperial/TP/95-96/57}}{\hbox{hep-th/9607107}} {\hbox{
  }}} }
{\vbox{\centerline { Near-Extremal Black Hole Entropy}
\vskip4pt
\centerline{and Fluctuating 3-Branes
 }
}}
\vskip -20 true pt
\medskip
\medskip
\centerline  { {  I.R. Klebanov\footnote {$^*$} {e-mail address:
 klebanov@puhep1.princeton.edu} }}

 \smallskip \smallskip

\centerline{\it Joseph Henry Laboratories }
\smallskip

\centerline{\it   Princeton University, Princeton, NJ 08544 }
\medskip
\centerline {and}

\medskip
\centerline{   A.A. Tseytlin\footnote{$^{\star}$}{\baselineskip8pt
e-mail address: tseytlin@ic.ac.uk}\footnote{$^{\dagger}$}{\baselineskip8pt
On leave  from Lebedev  Physics
Institute, Moscow.} }

\smallskip\smallskip
\centerline {\it  Theoretical Physics Group, Blackett Laboratory}
\smallskip

\centerline {\it  Imperial College,  London SW7 2BZ, U.K. }

\bigskip
\centerline {\bf Abstract}
\medskip
\baselineskip10pt
\noindent
We discuss the known microscopic interpretations of the
Bekenstein-Hawking entropy for
 configurations of intersecting M-branes.
In some cases the entropy scales as that of a massless field theory
on the intersection. A different situation, found for
configurations which reduce to 1-charge $D=5$ black holes
or 2-charge $D=4$ black holes, is explained by a gas of non-critical
strings at their Hagedorn temperature. 
We further suggest that the entropy of configurations reducing 
to 1-charge $D=4$ black holes is due to 3-branes moving within
5-branes.

\medskip
\Date {July 1996}
\noblackbox
\baselineskip 13pt plus 2pt minus 2pt
\lr\MalLen{J. Maldacena and L. Susskind, hep-th/9604042.}
\lr\DasMat{S. Das and S. Mathur, hep-th/9601152.}
\lr \dgh {A. Dabholkar, G.W. Gibbons, J. Harvey and {}F. Ruiz Ruiz,  \np
B340 (1990) 33;
A. Dabholkar and  J. Harvey,  \prl
63 (1989) 478.
}
\lr\mon{J.P. Gauntlett, J.A. Harvey and J.T. Liu, \np B409 (1993) 363.}
\lr\chs{C.G. Callan, J.A. Harvey and A. Strominger, 
\np { B359 } (1991)  611.}

\lr \CM{ C.G. Callan and  J.M.  Maldacena, 
  hep-th/9602043.} 
\lr\SV {A. Strominger and C. Vafa,   hep-th/9601029.}

\lr\MV {J.C. Breckenridge, R.C. Myers, A.W. Peet  and C. Vafa, HUTP-96-A005,  hep-th/9602065.}
\lr\vijay{V. Balasubramanian and {}F. Larsen, hep-th/9604189.}
\lr \CT{M. Cveti\v c and  A.A.  Tseytlin, 
\pl { B366} (1996) 95, hep-th/9510097. 
}
\lr \CTT{M. Cveti\v c and  A.A.  Tseytlin, \pr D53 (1996) 5619, 
 hep-th/9512031. 
}
\lr\LW{ {}F. Larsen  and {}F. Wilczek, 
  hep-th/9511064.    }
\lr\TT{A.A. Tseytlin, \mpl A11 (1996) 689,  hep-th/9601177.}
\lr \HT{ G.T. Horowitz and A.A. Tseytlin,  \pr { D51} (1995) 
2896, hep-th/9409021.}
\lr\khu{R. Khuri, \np B387 (1992) 315; \pl B294 (1992) 325.}
\lr\CY{M. Cveti\v c and D. Youm,
 UPR-0672-T, hep-th/9507090; UPR-0675-T, hep-th/9508058; 
  \pl { B359} (1995) 87, 
hep-th/9507160.}

\lr \jons{J.H. Schwarz, hep-th/9510086. }
\lr\ght{G.W. Gibbons, G.T. Horowitz and P.K. Townsend, \cqg 12 (1995) 297,
hep-th/9410073.}
\lr\dul{M.J. Duff and J.X. Lu, \np B416 (1994) 301, hep-th/9306052. }
\lr\hst {G.T. Horowitz and A. Strominger, hep-th/9602051.}
\lr\dull{M.J. Duff and J.X. Lu, \pl B273 (1991) 409. }
\lr \guv{R. G\"uven, \pl B276 (1992) 49. }
\lr \gups {S.S. Gubser, I.R. Klebanov and A.W. Peet, 
hep-th/9602135.}
\lr \gupse {S.S. Gubser, A. Hashimoto,  I.R. Klebanov and J.M.  Maldacena,
hep-th/9601057.}
\lr\juan{J. Maldacena, hep-th/9605016.}

\lr \dus { M.J. Duff and  K.S. Stelle, \pl B253 (1991) 113.}

\lr\hos{G.T.~Horowitz and A.~Strominger, Nucl. Phys. { B360}
(1991) 197.}
\lr\teit{R. Nepomechi, \pr D31 (1985) 1921; C. Teitelboim, \pl B167 (1986) 69.}
\lr \duf { M.J. Duff, P.S. Howe, T. Inami and K.S. Stelle, 
\pl B191 (1987) 70. }
\lr\duh {A. Dabholkar and J.A. Harvey, \prl { 63} (1989) 478;
 A. Dabholkar, G.W.   Gibbons, J.A.   Harvey  and {}F. Ruiz-Ruiz,
\np { B340} (1990) 33. }
\lr\mina{M.J. Duff, J.T. Liu and R. Minasian, 
\np B452 (1995) 261, hep-th/9506126.}
\lr\dvv{R. Dijkgraaf, E. Verlinde and H. Verlinde, hep-th/9603126;
hep-th/9604055.}
\lr\gibb{G.W. Gibbons and P.K. Townsend, \prl  71
(1993) 3754, hep-th/9307049.}
\lr\kap{D. Kaplan and J. Michelson, hep-th/9510053.}
\lr\hult{
C.M. Hull and P.K. Townsend, Nucl. Phys. { B438} (1995) 109;
P.K. Townsend, Phys. Lett. {B350} (1995) 184;
E. Witten, \np B443 (1995) 85; 
J.H. Schwarz,  \pl B367 (1996) 97, hep-th/9510086, hep-th/9601077;
P.K. Townsend, hep-th/9507048;
M.J. Duff, J.T. Liu and R. Minasian, 
\np B452 (1995) 261, hep-th/9506126; 
K. Becker, M. Becker and A. Strominger, Nucl. Phys. { B456} (1995) 130;
I. Bars and S. Yankielowicz, hep-th/9511098;
P. Ho{\v r}ava and E. Witten, Nucl. Phys. { B460} (1996) 506;
E. Witten, hep-th/9512219.}
\lr\Witten{E. Witten, hep-th/9507121.}
\lr\beck{
K. Becker and  M. Becker, hep-th/9602071.}
\lr\aar{
O. Aharony, J. Sonnenschein and S. Yankielowicz, hep-th/9603009.}
\lr\ald{{}F. Aldabe, hep-th/9603183.}
\lr\ast{A. Strominger, hep-th/9512059.}
\lr\ttt{P.K. Townsend, hep-th/9512062.}
\lr\pt{G. Papadopoulos and P.K. Townsend, hep-th/9603087.}
\lr\jch{J. Polchinski, S. Chaudhuri and C.V. Johnson, 
hep-th/9602052.}
\lr \ddd{E. Witten, hep-th/9510135;
M. Bershadsky, C. Vafa and V. Sadov, hep-th/9510225;
A. Sen, hep-th/9510229, hep-th/9511026;
C. Vafa, hep-th/9511088;
M. Douglas, hep-th/9512077. }

\lr \gig{G.W. Gibbons, M.J. Green and M.J. Perry, 
hep-th/9511080.}

\lr \dufe{M.J. Duff, S.  {}Ferrara, R.R. Khuri and 
J. Rahmfeld, \pl B356 (1995) 479,  hep-th/9506057.}

\lr\stp{H. L\" u, C.N. Pope, E. Sezgin and K.S. Stelle, \np B276 (1995)  669, hep-th/9508042.}
\lr \duff { M.J. Duff and J.X. Lu, \np B354 (1991) 141. } 
\lr \pol { J. Polchinski, \prl 75 (1995) 4724,  hep-th/9510017.} 
\lr \iz { J.M. Izquierdo, N.D. Lambert, G. Papadopoulos and 
P.K. Townsend,  \np B460 (1996) 560, hep-th/9508177. }

\lr \US{M. Cveti\v c and  A.A.  Tseytlin, 
\pl {B366} (1996) 95, hep-th/9510097.  
}
\lr\mast{J.M. Maldacena and A. Strominger, hep-th/9603060.}
\lr \CY{M. Cveti\v c and D. Youm,
 \pr D53 (1996) 584, hep-th/9507090.  }
 \lr\kall{R. Kallosh, A. Linde, T. Ort\' in, A. Peet and A. van Proeyen, \pr { D}46 (1992) 5278.} 
\lr \grop{R. Sorkin, Phys. Rev. Lett. { 51 } (1983) 87;
D. Gross and M. Perry, Nucl. Phys. { B226} (1983) 29. 
}
\lr \myers{C. Johnson, R. Khuri and R. Myers, hep-th/9603061.} 

\lr \myk{R.R. Khuri and R.C. Myers, hep-th/9512061.}

\lr\lup {H. L\" u and C.N. Pope, hep-th/9512012; hep-th/9512153.}

\lr \kt{I.R. Klebanov and A.A. Tseytlin, hep-th/9604089.}

\lr \CS{M. Cveti\v c and  A. Sen, unpublished.}
\lr \at{ A.A. Tseytlin, hep-th/9604035.}
\lr \pap{G. Papadopoulos, hep-th/9604068.}

\lr \green{M.B. Green and M. Gutperle, hep-th/9604091.}
\lr \sen{A. Sen, \mpl  { A10} (1995) 2081, 
 hep-th/9504147. }
\lr \gib{G.W.  Gibbons and K. Maeda, \np {B298} (1988) 741.} 
\lr \dabb{  A. Dabholkar, J.P. Gauntlett, J.A. Harvey and D. Waldram, 
  hep-th/9511053.   }

\lr\mina{M.J. Duff, J.T. Liu and R. Minasian, 
\np B452 (1995) 261, hep-th/9506126.}

\lr\polch{
J.~Polchinski, Phys. Rev. Lett. 75 (1995)
4724,  hepth/9510017.}

\lr \CT {M. Cveti\v c and  A.A.  Tseytlin, hep-th/9606033.}
\lr \KT{I.R. Klebanov and A.A. Tseytlin, hep-th/9604166.}
\lr\HMS {G.T. Horowitz, J.M.  Maldacena and A. Strominger,  
hep-th/9603109.}

\lr\HLM{G.T. Horowitz, D.A. Lowe and J.M. Maldacena, 
hep-th/9603195.}

\lr\susm{J.M.  Maldacena and L. Susskind, hep-th/9604042.}
\lr\HSS{G.T. Horowitz  and A. Strominger, 
Nucl. Phys.  { B360} (1991) 197.}

\lr\HK{A.~Hanany and I.R.~Klebanov, hep-th/9606136.    }
\lr\mald{ J.M.~Maldacena, hep-th/9605016.   }

\lr \CYY{M. Cveti\v c and D. Youm, hep-th/9508058; hep-th/9512127;  hep-th/9603100;   hep-th/9605051.} 

\lr \Peet{A.W. Peet, hep-th/9506200.}

\lr\ortin{
E.~Alvarez and T.~Ortin, \mpl { A7} (1992) 2889.}

\lr \byts{B. Harms and Y. Leblanc, \pr D47 (1993) 2438;
A.A.~Bytsenko, K.~Kirsten and S.~Zerbini,
\pl {B304} (1993) 235.} 

\lr \pt{
G.~Papadopoulos and P.K.~Townsend,
hep-th/9603087. }

\lr \at{
A.A.~Tseytlin, 
hep-th/9604035.}
\lr\gkt{J.P.~Gauntlett, D.A.~Kastor and J.~Traschen, 
hep-th/9604179.}

\lr \town{P.K. Townsend, \pl B350 (1995) 184, hep-th/9501068.}

\lr \dvv{
R. Dijkgraaf, E. Verlinde and H. Verlinde, hep-th/9603126;
hep-th/960726.}
\lr\pop{
N. Khviengia, Z. Khviengia, H. L\"u  and  C.N. Pope, 
  hep-th/9605077.}
\lr\ds{M.J. Duff and K.S. Stelle, \pl B253 (1991) 113.    }
\lr\guv{ R. G\" uven, \pl  {B276} (1992) 49.}
\lr \polc{J. Hughes, J. Liu and J. Polchinski, \pl B180 (1986) 370.}
\lr\john{
J.H. Schwarz,  \pl B367 (1996) 97, hep-th/9510086.}
\lr\ast{A. Strominger, hep-th/9512059;
P.K. Townsend, hep-th/9512062;
K. Becker and  M. Becker, hep-th/9602071;
O. Aharony, J. Sonnenschein and S. Yankielowicz, hep-th/9603009.}

\lr\beh{ K. Behrndt, E. Bergshoeff and B. Janssen, 
hep-th/9604168.}
\lr\kap{C. Callan, J. Harvey and A. Strominger, \np B359 (1991) 611;
G.W. Gibbons and P.K. Townsend, \prl  71
(1993) 3754, hep-th/9307049;
D. Kaplan and J. Michelson,  \pr D53 (1996) 3474, hep-th/9510053.}

\lr\ach{A. Achucarro, J. Evans, P.K. Townsend and D. Wiltshire, \pl B198 (1987) 441.}
\lr\moo{M.B. Green, J.A. Harvey and G. Moore, hep-th/9605033.}

\newsec{Introduction}
The 11-dimensional M-theory  is expected to provide a unified description
of different string theories. In particular,  classical
solutions of 11-dimensional supergravity unify various R-R and NS-NS
charged soliton solutions of  type IIA $D=10$ string theory.
Supersymmetric black holes in $D <10$ have been discussed
from the 11-dimensional perspective in several recent papers \refs{\pt,\at,\KT,\CT,\HK}.
The aim of such discussions  is twofold: 
(1) to find ways of understanding 
the Bekenstein-Hawking entropy in statistical terms 
using hints  about the relevant degrees of freedom 
provided by the  representation of the  black holes as  
intersecting M-branes; 
(2) to obtain information about the quantum M-theory
(which virtual configurations  \ast\ are relevant, etc.)
in the process of  explaining the  thermodynamic  properties of the 
classical solutions.

The knowledge of the structure of possible  stable classical  configurations 
of M-theory 
has recently been improved substantially through
the construction of composite solutions representing 
intersecting M-branes \refs{\guv,\pt,\at,\kt,\gkt,\pop}. 
 This is  an important step: 
just like the basic 2-brane \refs{\ds,\guv} and 5-brane \refs{\guv},
these intersecting (or bound) objects
seem to play a  fundamental  role in  the theory. 
While the basic M-branes preserve 1/2 of the supersymmetries, the
intersecting configurations can preserve only $1/2^N$ where 
$N=2, 3$ or $4$.
Upon compactification the intersecting M-branes
describe the most general black holes. Of particular interest
from the point of view of  entropy
are the supersymmetric black holes with regular horizons
in $D=4$ or 5.
The representation of 
black holes by compactified intersecting M-branes
has the advantage of being symmetric in different charges.
The counting of states directly in the M-theory should
provide a unified picture of the statistical description of
the R-R charged (D-brane)  
and the NS-NS charged (conformal sigma-model) black holes 
found in string theory.
{}Furthermore, it may teach us something about the
M-theory itself.

In previous work \refs{\KT,\HK}
it was indeed found that the Bekenstein-Hawking
entropy  gives  important clues about 
the degrees of freedom of M-theory.
One such interesting development is the claim that the excitations
of a 5-brane with one compact transverse direction 
are strings on
its $D=6$ world volume \juan. 
These strings originate from 2-branes intersecting
the 5-brane along one dimension, with the other wrapped around
the circle. A similar phenomenon leading to strings in $D=4$
seems to be responsible for the entropy of two 5-branes intersecting
over a 3-brane \HK. As we discuss in more detail below, in this case
the string comes from a 5-brane compactified on $T^4$.
In this paper we propose something even more surprising: we argue
that the entropy of a 5-brane with two compact transverse directions
is explained by 3-branes moving inside it. These 3-branes are
5-branes compactified on $T^2$. 
The presence of 1-branes and 3-branes in M-theory hints at
a relation with the type IIB string (or {}F-) theory.


\newsec{Intersecting M-branes as black holes in various dimensions}
As discussed in \CT, the charged  $D <10$ black hole solutions 
of type II  string theory  compactified on a torus
 can be represented   as low-dimensional images of non-extremal generalization
of supersymmetric intersecting 
2- and 5-brane configurations  \refs{\pt,\at,\KT}  in $D=11$.
The latter can be viewed as  black anisotropic p-brane solutions
of 11-dimensional supergravity parametrized   
by  $N$ charges,
$Q_i$, and the `non-extremality' parameter $\m$.
 The interpretation of charges   depends on  specific composition
of M-branes:  $Q_i$ may be a 2-brane `electric' charge, a  
5-brane `magnetic' charge or
a momentum along  compact  direction,  i.e. Kaluza-Klein electric charge.

Sending
 $\m \to 0$ we obtain an extremal black hole with $N$ charges.
On the other hand, setting all charges to zero
gives the neutral Schwarzschild solution 
with mass $\sim \m$.  While a composite configuration 
of several non-extremal M-branes is not described by a static solution,
the non-extremal version of supersymmetric intersecting M-brane  backgrounds
has a  simple form which 
directly generalizes the `harmonic function product' structure
of extremal  solutions (in the metric one is only to include the 
factors of Schwarzschild function $f =1 - {2\m \ov r^{D-3}}$,
where $D= 11-p$, and to change the  parameters 
of harmonic functions from $Q_i$ to
$\Q_i = \sqrt {Q_i^2 + \m^2} -\m $).\foot{Our notation 
differs slightly from 
that in \CT: $\m$ is rescaled by factor of 2 and the number of charges 
is now denoted by $N$, not $n$.}
The $\m$-extended solution,
which interpolates between the
extremal and the Schwarzschild solutions, thus 
preserves certain simple features  of the extremal 
solution. 
This suggests that $\m$  may  be viewed as 
a parameter of `soft' supersymmetry breaking.
The idea  is  then to achieve some understanding of  
the properties of the physically relevant Schwarzschild  black hole 
by expanding  near the supersymmetric extremal point, 
i.e. by doing perturbation theory in $\m$. 
\subsec{Review}
Let $D=11-p$ denote the number of non-compact dimensions
of intersecting M-brane configuration with $p$ being the  total number of 
(compact) internal coordinates $y_n$. Then (some of) 
the relevant  M-brane configurations 
which upon dimensional reduction  represent  black holes  with $N$ charges 
in $3<D <11$ are

$D=10:  \ \  N=1: \  0\ua $

$D=9:  \ \ \ N=1: \  2 ; \ \ \  \  N=2:  \ 2\ua  $

$D=8:  \ \ \ N=1: \  2_1 ; \ \ \   N=2:  \ 2_1\ua  $

$D=7:  \ \  \ N=1: \  2_2 ; \ \ \   N=2:  \ 2\bot 2  $

$D=6:  \ \ \  N=1: \  2_3, \ 5  ; \ \ \ \   N=2:  \ 5\ua  $

$D=5:  \ \  \ N=1: \  2_4, \ 5_1 ; \ \ \   N=2:  \ 2\bot 5;  \ \ \   N=3: 
\ 2\bot 2\bot 2 , \  \ 2\bot 5\ua    $

$D=4:  \ \  \ N=1: \  2_5,\ 5_2 ; \ \ \  
 N=2: \  5\bot 5 ; \ \ \ 
N=3:  \ 2\bot2\bot  5, \ 2\bot 5\bot  5, \ 5\bot 5 \bot  5; $ 

$  \ \ \ \  \ \ \ \ \ \ \ \ \ \ 
 N=4:  \    2\bot2\bot  5\bot  5, \ 5\bot 5 \bot  5\ua $

\noindent
Here $2\bot 2$  stands for the orthogonal intersection of two  2-branes
over a point, $2\bot 5$ -- for the intersection of a 2-brane and a 5-brane 
over a 1-brane, $5\bot 5$ -- for intersection of  two  5-branes
over a 3-brane,  and $\ua$ denotes a momentum (boost) along 
a compact direction 
which is possible to add when there is a null isometry in the extremal limit.
{}For completeness we have included  the  $D=10$ black hole ($0$-brane)   
which  does not  admit an M-brane representation
but may be thought of  as a  reduction of the 
boosted Schwarzschild black hole in $D=11$.
The subscripts indicate the number of compact transverse coordinates. 
{}For example,  $2_2$ stands for the 
$2$-brane solution which depends only on 6 of the 8 
transverse coordinates. In other words, this is a 2-brane
``averaged'' over two compact transverse directions.
This is equivalent to a special 
case of $2\bot 2$ where the charge of the second 2-brane set equal to zero, 
i.e. $2\bot 2^{(0)}$.
Similarly, $2_4= 2\bot 5^{(0)} = 2\bot 2^{(0)}\bot 2^{(0)} $ and 
$5_1= 5\bot 2^{(0)}, \ 5_2= 5\bot 5^{(0)}, $  etc. 

The  remarkable feature of unboosted configurations is that the metric is diagonal and thus all charges are contained in the antisymmetric 3-tensor field strength. Let us note  that there exist  also other embeddings
of $D=4$  black holes into 11-dimensional theory which involve Kaluza-Klein monopole (6-brane in $D=10$) \refs{\town,\at}.

The Einstein-frame metric of the corresponding  black holes in $D=10, ..., 4$
dimensions  has the following universal form \CT\foot{Particular cases
of equivalent  non-extremal  black hole solutions were  constructed in 
\refs{\HSS,\Peet,\CYY,\HMS}.}
\eqn\met{
ds^2_D=   h^{1\ov D-2} (r)  \big[ - h\inv (r) f(r)  dt^2+   
f\inv (r) dr^2+r^2d\Omega_{D-2} \big]
\  ,}
$$
h(r) = \prod^N_{i=1}  H_i(r) \ , \ \ \ \  f(r) = 1 -  { 2\mu \ov r^{D-3} } \ ,
\ \ \ \ 
H_i(r) = 1 + { \Q_i \ov r^{D-3} }  \ ,  
\ \   
\ \  \Q_i = \sqrt {Q_i^2 + \m^2} -\m \ . $$
The corresponding  ADM mass and 
the Bekenstein-Hawking entropy  are given by \CT\
\eqn\mass{
 M  = b ( \sum^N_{i=1} 
\sqrt {Q_i^2 + \m^2}  +  2 \l  \m ) \ ,  \  }
\eqn\entr{
S= {2\pi A_9\ov \k^2} =  2b \  {  \m  \ov   T_H }
=  c (2\m)^\l   {  \prod_{i=1}^N \sqrt{  \sqrt {Q_i^2 + \m^2 } + \m }
}   \ , }
$$ b \equiv 
{\omega_{D-2}\ov 2\k^2 } (D-3) V_p \ , \ \  \ \ \ \ \ \
 c\equiv {2\pi\omega_{D-2} \ov \k^2} V_p= {4\pi b\ov D-3}  \ . $$
Here $V_p= L_1 ... L_p $  ($p=11-D$) is 
the volume of the compact internal space, 
 $\k^2 (= V_p  \k^2_D)$  is the gravitational constant in 11 dimensions, 
$T_H$ is the Hawking temperature and $\l$ is the 
important scaling  index \kt,      
\eqn\lam{
\l \equiv {D-2\ov D-3  } -  {N\over 2}  \ . }
The expression for $S(\m, Q_i)$ interpolates between the 
standard Schwarzschild  entropy, $S(\m,Q_i=0) \sim \m^{D-2\ov D-3}$,
and the entropy of extremal black holes, $S(\m=0,Q_i)$.

The mass   and  the entropy  satisfy the following relation 
\eqn\rel{
 {    \partial M \over  \partial \mu } = 
2b {  \partial  \ln  S \over  \partial \ln  \mu}   \ , \ \ \ 
i.e. \ \ \ \ 
{  T_H} {  \partial  S \over  \partial  \mu}   
=     {    \partial M \over  \partial \mu } \ . }
This thermodynamic equation  
(related to the `first law'  $ T_H dS(\m, Q_i)  
= dM (\m, Q_i) - \sum^N_{i=1} \Phi_i dQ_i$)
 is  valid for all $\m$ and $Q_i$, ranging from the  
Schwarzschild  ($Q_i=0$) case all the way to 
the extremal limit ($\m=0$). 
This explains why 
the proportionality  between the entropy and the area
of the horizon can be assumed
to be true also in the extremal  limit: it 
 certainly holds  in the non-extremal (or near-extremal) case 
and thus should be meaningful
 also in the limit  $\m \to 0$.\foot{The use of 
$\m$ as a `regularisation parameter' to define
the entropy in the extremal limit resolves also  the following paradox.
Starting with an extremal  solution parametrized by an arbitrary 
harmonic function, one may deform it to a multicenter configuration
which has the same mass, but  always zero entropy (only the
single-center solution may have a non-zero entropy).
Thus, even an infinitesimal transverse
separation of M-branes leads to a change of entropy
at no energy cost. The point, however, is that 
the `separated' configuration  {\it does not }
have a stable non-extremal generalization and, thus, its entropy
is not well-defined.}

The entropy  has a non-zero  $\m\to 0$   limit only if $\l=0$ \kt, 
i.e. for the black holes with  $D=5,N=3$  and $D=4,N=4$
which have non-singular horizons in the extremal  limit.
In these cases
\eqn\yyy{
M =  b  \sum^N_{i=1} 
\sqrt {Q_i^2 + \m ^2}  \ , 
\ \ \ \ \ \  S=   c
 \prod_{i=1}^N \sqrt{  \sqrt {Q_i^2 + \m^2}  
+ \m } \ .  }
Using the $D=11$ charge quantization conditions one 
can further  show that, for all intersecting M-brane configurations
with $D=5,N=3$  and $D=4,N=4$ one has \KT\
\eqn\een{
S_0= (S)_{\m=0} = c
 \prod_{i=1}^N \sqrt{ Q_i} =   2\pi   \prod_{i=1}^N  \sqrt {n_i} \ , }
where $n_i$ are the integer values of the quantized charges.
A statistical interpretation
of this entropy  based on the existence of the 
corresponding intersecting M-brane solutions
$2\bot 5 \ua$ and $5\bot 5 \bot 5 \ua$
was discussed in \KT.
These configurations are characterized by
a common intersection string on a 5-brane (i.e. $c_{eff} =6, \  d=6$ string)
with  momentum along it 
as one quantum number, and the effective winding number proportional 
to $n_1 n_2$ in the $D=5$ case, and $n_1n_2n_3$  in the $D=4$  case.

This paper is primarily devoted to
a statistical interpretation of the {\it near-extremal}   terms in $S$
suggested by the intersecting M-brane representations.

\subsec{Sub-leading entropy in the regular cases ($\l=0$) }
{}For solutions which have regular horizons,
the $D=5,N=3$  and $D=4,N=4$ cases, it is interesting to study the
leading near extremal corrections to the mass and Bekenstein-Hawking
entropy.
As follows from \yyy, 
\eqn\yyyw{
M =  b [ \sum^N_{i=1} Q_i  + \ha  \m ^2 \sum^N_{i=1} Q\inv_i  + O(\m^4)] 
= M_0 + E + O(\m^4) \ , }
\eqn\yw{  S=  c
 \prod_{i=1}^N \sqrt{ Q_i}  [ 1  + \ha \m  \sum^N_{j=1} Q\inv_j + O(\m^2)]
= S_0 + \Delta S + O(\m^2) \ , }
\eqn\uuu{
\Delta S =   {c\ov \sqrt {2b}}\bigg[ \sum^N_{i=1} 
\prod^N_{j\not=i =1} Q_j\bigg]^{1/2}  \  \sqrt E \ . }
The dependence of $\Delta S$ on $E$ is characteristic of
a $1+1$-dimensional field theory, but the dependence of the
prefactor on $Q_i$ needs an explanation.
In \refs{\HMS, \HLM} an explanation was proposed based on the presence of
antibranes in $D=10$ string theory picture. 
The antibranes seem to play a role similar to the left-movers
on a string with mostly  right-moving modes. As the number of antibranes
(or left-movers) is sent to zero, the configuration returns to extremality.
In the special case of equal charges  
(the  Reissner-Nordsr\"om background with constant  scalar fields) \uuu\ 
gives 
\eqn\yyw{
{\Delta S\ov  S_0} = {N} \sqrt { E\ov 2 M_0} \ . }
The factor of $N$ is explained by antibranes giving contributions equal
to that of the left-movers. There exists a special case, however, where
the antibranes are suppressed. 
{}For the $5\bot 5\bot 5\ua$ configuration
this happens if we choose
\eqn\sup{\mu\ll Q_4 \ll Q_1, Q_2, Q_3  \ ,  }
where $Q_4$ is the Kaluza-Klein charge.
Now the correction to the entropy is
\eqn\corr{
\Delta S = {c\over\sqrt{2b} }\sqrt {Q_1 Q_2 Q_3 E } \ . }
The charges $Q_i$  have  the following
values \KT,
\eqn\charges{
Q_1 = {n_1\over  L_6 L_7} ({\kappa\over 4\pi})^{2/3}
 ,\ \ 
Q_2 = {n_2\over  L_4 L_5} ({\kappa\over 4\pi})^{2/3}
, \ \ 
Q_3 = {n_3\over  L_2 L_3} ({\kappa\over 4\pi})^{2/3}
 , }
where $L_i$ is the length of the $i$-th  compact dimension.
The left- and right-moving momenta are given by
\eqn\tyu{  P_L= {2\pi N_L\over L}\ ,\qquad
P_R= {2\pi N_R\over L}\ . }
Near extremality we have 
\eqn\ipo{ N_R= {\cal N} +n\ ,\qquad  N_L=n\ , \qquad (n\ll {\cal N})\ , 
 \ \ \ \ i.e. \ \  E ={4\pi n\ov L} \ . } 
In terms of the integer charges, we find
\eqn\correct{
\Delta S =2 \pi
\sqrt{n_1 n_2 n_3 n} \ . }
As explained in section 3, this is the correct relation for a 
$1+1$-dimensional theory on a circle of effective length 
$n_1 n_2 n_3 L_1$ with central charge $c_{eff}=6$.
In fact, the general entropy formula is
\eqn\genent{
S =2 \pi \sqrt{
n_1 n_2 n_3}
\left (\sqrt {N_L } + \sqrt {N_R}  \right ) \ . }
In section 3 we further check the consistency of this formula.




\subsec{ Near-extremal entropy for $\l\not=0$  }
As follows from \entr,\lam\ and \yyy, setting a charge to zero 
(i.e. going from a configuration with $N$ charges to 
one with $N-1$ charges) supplies a factor of $\m$ in the entropy.
Thus, for $\l\not=0$  we find, in the near-extremal limit,
\eqn\ii{
M = M_0 +  E +  O(\m^2) \ , \ \ \ \ \  \ 
E=  2 b \l \m      \ ,  }
\eqn\pop{
 S= c_1 \prod_{i=1}^N \sqrt  { Q_i }\  E^\l    \ , 
\ \ \ \  \  \ \ \ \  c_1= c(b\l)^{-\l} \ .  }
In terms of  the Hawking temperature $T$ 
(the leading near-extremal  term  in  $T_H$  in \entr)
 satisfying  $dE= TdS$,
as implied by  \rel,  we find  
\eqn\sca{ S= c_2 \prod_{i=1}^N ({Q_i})^{1\ov 2(1-\l)} \  T^{\l\ov 1-\l}  
 \  , \ \ \ \ \ \ \ c_2  = (c b^{-\l})^{1\ov 1-\l} \ . }
Since the power of $T$ is not positive  for $\l >  1$,  the   canonical ensemble description applies only for 
$\l <1$ while for $\l \geq 1$ one should use the  microcanonical one.

Let us recall that the entropy of a
massless ideal gas in $p$ spatial dimensions scales as
$S_p \sim E^{p\ov p+1}$.
In \refs{\gups,\kt} it was noted that there are cases where
$\l={p\ov p+1}$, so that the Bekenstein-Hawking entropy
may be interpreted as being due to massless fields on the
$p$-brane.
Here is a list of such cases:

(1) $D=9,N=1$, i.e. $p=2, \l={2\ov 3}$. This is the basic
2-brane of M-theory,  for which  
$S \sim \sqrt Q E^{{2\ov 3}} \sim Q^{{3 \ov 2}} T^2$ \kt;

(2) $D=7,N=1$, i.e. $p=4, \l={3 \ov 4}$,  
which is represented by the $2_2= 2\bot 2^{(0)}$ configuration.
In the type IIB theory this is represented by the self-dual 
3-brane in 10 dimensions.
Indeed, dimensionally reducing $2_2$ to $D=10$ along one 
of two compact transverse directions 
we get a $2_1$-brane of type IIA theory, which is T-dual 
to the 3-brane of type IIB.\foot{More generally,  $2\bot 2$ gives upon reduction to $D=10$
$2\bot 1$ solution of type IIA theory -- fundamental string intersecting R-R 2-brane. $T$-duality along the string direction converts this into 3-brane of type IIB with extra momentum along one direction
equal to string winding number or charge of the second 2-brane in 
$2\bot 2$. Setting this charge to zero gives unboosted 3-brane, which is indeed $T$-dual to  $2_1$-brane in $D=10$.} 
Here one finds the following scaling \gups,
$S \sim \sqrt Q  E^{{3 \ov 4}} \sim Q^2 T^3$. 
 
(3) $D=6,N=1$, i.e. $p=5, \l={5 \ov  6}$.  
This is the basic 5-brane of M-theory, with the scaling  
$S \sim \sqrt Q E^{{5 \ov  6}} \sim Q^3 T^5$ \kt.

The feature shared by  the cases (1)--(3) is that,
for the extremal solution, the coupling 
strength  is well-behaved at the location of the brane ($x=0$).
Although the scaling of $S$ in these cases is natural,
the precise number of degrees of freedom is still awaiting a complete
explanation \refs{\gups,\kt}.
{}For other $D >5$  black holes 
 (including the $D=10$ one  \gupse\ with $\l={\textstyle{9\ov 14}}$)
even the scaling
exponents defy a simple explanation in terms of a free gas of massless
particles. Perhaps the interactions or the massive string
modes need to be taken into account.

In this paper we concentrate on the $D=5,4$ black holes 
with fewer than the critical  ($N_*=3,4$ respectively) number of charges. 
If the number of charges equals $N_*-1$, then  $\l=\ha$ which leads to
\eqn\uu{ S {(\l=\ha)}
\sim  \prod_{i=1}^N \sqrt  { Q_i } 
\sqrt E \sim  \prod_{i=1}^N  { Q_i }   T \ , }
i.e. the scaling behavior characteristic of a $1+1$-dimensional
field theory. This is 
not surprising since the M-brane configurations
corresponding to the $D=5,N=2$ and $D=4, N=3$ cases,  
$2\bot 5$ and $5\bot 5\bot 5$ respectively, intersect over
a common string.
The  entropy can then be attributed to massless
modes on the string, whose effective winding number
is  proportional to 
$\prod_{i=1}^N  { Q_i }$. A detailed discussion of the
$5\bot 5\bot 5$ case is provided in section 3.1.

In the cases where the number of charges is  $N_*-2$
we get $\l=1$, i.e.
\eqn\vv{
S {(\l=1)}=  { E\ov T} \sim  \prod_{i=1}^N \sqrt  { Q_i }  E   
\ , \ \ \ \ \   \   T=\const.  }
The corresponding configurations are
$5_1= 5 \bot 2^{(0)}$ or $2_4$ (which we discuss in section 3.1)
for $D=5$, and $5\bot 5$  for $D=4$. 
This is the same scaling as in  the 
case of the $D=10$ 5-brane \refs{\kt,\juan}, to which 
$5_1$ indeed reduces upon compactification on a circle.
{}For $5_1$ the entropy is  naturally explained by 
non-critical strings in $5+1$ dimensions \juan, for
$2_4$ -- in $2+1$ dimensions, while
for $5\bot 5$ -- by non-critical strings on the
$3+1$-dimensional intersection \HK.
The latter explanations are discussed in section 3.1.

{}Finally, consider the case 
where the number of charges is $N_*-3$.
This is possible only for the $D=4, N=1$ black holes, and 
the two relevant M-brane configurations are
$2_5$ and $5_2$.
Here we find 
$\l={3 \ov 2}$, i.e.  
\eqn\tre{
S {(\l={\textstyle {3\ov 2}} )} \sim   
\sqrt { Q }  E^{{3 \ov 2}} \sim     Q\inv T^{-3}   \ . }
In section 3.2 we argue that this scaling is explained by dynamical
{\it 3-branes} in $5+1$ dimensions.

\newsec{The Entropy of Intersecting 5-Branes } 
We shall  now  focus on the intersecting M-brane configurations
which, upon dimensional reduction, correspond to the
$D=4$ black holes with $N=$3, 2, or 1 charges. 
{}For each of these cases we suggest a separate microscopic explanation of
the near-extremal Bekenstein-Hawking entropy.

\subsec{Entropy due to non-critical strings}

Let us start with the intersecting M-branes which, upon wrapping them
over a $T^7$, reduce to $D=4$ black holes with 3 charges.
The example that we shall discuss in detail is $5\bot 5\bot 5$, 
 involving  $n_1$ 5-branes 
positioned in the $(12345)$ hyperplane, $n_2$ 5-branes
positioned in the $(12367)$ hyperplane, and $n_3$ 5-branes
positioned in the $(14567)$ hyperplane.\foot{The other
possibilities where some of the 5-branes are replaced by the
2-branes are related to this by the U-duality of the $D=4$ theory.}
The classical solution describing this configuration with the
non-extremality parameter $\mu$ is given by \CT
\eqn\sol{
ds^2_{11} = (H_1 H_2 H_3)^{2/3}\big [ (H_1 H_2 H_3)\inv  (-f dt^2+ dy_1^2)+ 
(H_1 H_2)\inv  (dy_2^2 + dy_3^2) }
$$
+( H_1 H_3)\inv  (dy_4^2+ dy_5^2) + (H_2 H_3)\inv  (dy_6^2+ dy_7^2)+ 
f^{-1} dr^2 + r^2 d\Omega_2^2 \big ] \ , $$
$$
H_i =  1+ {\Q_i\over r}\ ,
\qquad f=1-{2\mu\over r}\ , \ \ \ \ \Q_i=\sqrt{Q_i^2 +\m^2} -\m \ .   $$
The near-extremal Bekenstein-Hawking entropy of this solution  is given by
\pop,\uu
\eqn\rer{
S= {8\pi^{{3/2}}\over \kappa}
\left ( Q_1 Q_2 Q_3 \prod_{i=1}^7 L_i E \right )^{1/2} =
2\sqrt \pi
\sqrt{n_1 n_2 n_3 L_1 E} \ , }
where we have used \charges.
This has the same form as the entropy vs. energy of a $1+1$ dimensional
field theory with central charge $c_{eff}$ defined on a circle of
length $L_{eff}$,
\eqn\str{
S=  2  \sqrt{ \pi}
\sqrt {{\textstyle{1\ov 6}} c_{eff} L_{eff} E } \ . }
Thus, the near-extremal Bekenstein-Hawking entropy is explained by
massless modes on the intersection of the 5-branes.
One possible identification is that $c_{eff}=6$, while
$L_{eff}= n_1 n_2 n_3 L_1$. This is consistent with what was necessary
to explain the extremal entropy of the boosted $5\bot 5\bot 5$
solution \KT. Note also that \rer\ agrees with the general formula
\genent\ after we substitute $E=4\pi N_L/L_1= 4\pi N_R/L_1$.

In \KT\ it was proposed that the massless
degrees of freedom come from collapsed 2-branes with triple boundaries,
one boundary lying in each of the 5-brane hyperplanes.
Such 2-branes provide triple connections of the 5-branes near the
intersection string. An argument analogous to that in \susm\ 
explains why the effective length  of circle 
on which  such degrees of freedom  move is
enhanced by the factor $n_1 n_2 n_3$.

Let us now set $n_3=0$ (i.e.  $\Q_3=0$,
 $H_3=1$)  in \sol. The resulting solution 
describes the $5\bot 5$ configuration which, upon compactification,
reduces to the $D=4$ black hole with 2 magnetic charges,
$n_1$ and $n_2$. As shown in \HK, its near-extremal entropy
reveals an interesting connection with non-critical strings
propagating on the $3+1$-dimensional intersection. Let us repeat the
essential points of the argument in \HK.

As a function of the energy, $E=M-M_0$, the near-extremal entropy is \pop,\vv
\eqn\bekstring{
S = 4\pi \sqrt{Q_1Q_2}\ E \ =     2^{7/6} \pi^{5/6}  \k^{2/3}
\big( {n_1 n_2 \ov L_4L_5L_6L_7} \big)^{1/2} \ E   \ .
}
This has the same form as the entropy of
a gas of strings at its Hagedorn temperature
(which is  dominated  by  the contribution of one long non-winding string)
\eqn\genstring{
S= 2\pi \sqrt{{\textstyle{1\ov 6}}c_{eff}\alpha'_{eff}}\ E \ . }
Here $c_{eff}$ is the central charge of the world sheet
degrees of freedom, while
$T_{1\ eff}={1\ov 2\pi \alpha'_{eff}}$ is the string tension.
To achieve agreement between \genstring\ and \bekstring,
we take \HK\  
\eqn\tension{ c_{eff}=6 \ , \ \ \ \ \ 
T_{1\ eff}= {L_4 L_5 L_6 L_7\over n_1 n_2}\ T_5 \ , }
where
\eqn\ttt{
T_5 =( {\pi\over 2})^{1/3} \kappa^{-4/3}
\ , }
is the tension of the 5-brane of M-theory \KT.
{}For $n_1=n_2=1$ the string tension equals that of a 5-brane
wrapped over the $T^4$ in the $(4567)$ directions.
{}For higher values of $n_1$ and $n_2$,
the tension of the wrapped 5-brane is reduced by a multiplicative factor.
A similar reduction of tension was
necessary for explaining the entropy of a D-string
moving within a number of parallel type IIB 5-branes \juan.
The value of $c_{eff}$ seems puzzling since, for a string in
$3+1$ dimensions, one would expect $c_{eff}=3$.
However, as we will show in the next paragraph, such a naive
expectation fails also for a string in $2+1$ dimensions -- one needs
an exact theory, such as the D-branes, to count the central charge.

As a slight digression, let us note that the 5-brane wrapped over $T^4$
also explains the near-extremal entropy of the $2_4$ configuration,
which reduces to a single-charge $D=5$ black hole. In this case,
the near-extremal entropy is related to energy according to \vv\  \CT
\eqn\fiveone{S= 2\pi \sqrt Q E\ , \qquad \ \  Q=  {n
\over  L_3 L_4 L_5 L_6} \big({\k \ov \sqrt 2 \pi} \big)^{4/3}\ ,}
where $3, 4, 5$ and $6$ are the compact directions transverse to
the $n$ 2-branes. Once again, the entropy has the form
\genstring\ characteristic of a string gas at Hagedorn temperature,
with 
\eqn\newval{ c_{eff}=6\ , \qquad 
T_{1\ eff}= {L_3 L_4 L_5 L_6\over n}\ T_5 \ .
}
Now the non-critical string originating from the wrapped 5-brane
moves in 
$2+1$ dimensions (within the 2-branes). {}From this point of 
view, the value of the central charge
necessary for matching the entropy seems mysterious.
However, upon reduction to 10 dimensions this value is confirmed
using the D-brane count (the resulting configuration, $2_3$,
is T-dual to the D5-brane considered in \juan).

As we have seen, the near-extremal entropy of 
the $5\bot 5$ configuration is  reproduced  by non-critical strings which
may be  interpreted as  5-branes wrapped around $T^4$. 
Presumably, this is related to the fact that there exists
an extremal $5\bot 5\bot 5$ configuration 
which preserves $1/8$ of the original 
supersymmetries \refs{\pt,\at,\gkt}:
we can add a 5-brane in the $(a4567)$
 hyperplane, with $a=1, 2$ or $3$.  {}From the point of
view of the extremal $5\bot 5$, this adds a long
straight string on the $3+1$ dimensional intersection.
We believe that the ``almost-supersymmetric'' objects responsible
for the near-extremal excitations of the $5\bot 5$ configuration
are large loops
of such string (if the string bends slowly, the supersymmetry
breaking is small). Unlike the completely straight strings,
such loops do not carry any charge (winding number).

Let us emphasize that $c_{eff}$ is equal to 6 both for the $5\bot 5\bot
5$
intersection string and for the
gas of non-critical strings on the
$5\bot 5$ intersection.
 This is necessary for consistency of our
interpretation because in both cases we are dealing with the same
non-critical string theory.\foot{It may be related 
to  the  string theory on the 5-brane  discussed in \dvv\
and in \juan.}
On the $5\bot 5$ intersection the strings
are free to wander around. Adding a large number of 5-branes in the
third hyperplane pins one long string down and makes it semiclassical,
so that the entropy is due to its small fluctuations.

\subsec{Entropy due to fluctuating 3-branes}

By analogy, we may now interpret the near-extremal entropy of
the solution \sol\  with $n_2=n_3=0$. This describes $n_1$
coincident 5-branes positioned 
 in the $(12345)$ hyperplane and 
``averaged''
over the compact directions $6$ and $7$, i.e. $5_2$.
Upon compactification on $T^7$, it reduces to the $D=4$ black hole
with one magnetic charge, $n_1$. We expect that the near-extremal
entropy of this solution is related to the existence of the
$5\bot 5$ configuration in the following sense: a 5-brane
in the $(abc67)$ hyperplane wrapped around the $(67)$ directions is
a straight 3-brane positioned inside the $n_1$ 5-branes
($a$, $b$ and $c$ are three orthogonal directions
within the $(12345)$ hyperplane). Clearly, the 3-brane has 
massless fermionic modes associated with the fact that it breaks $1/2$
of the supersymmetries present on the 5-brane, while
their bosonic partners
are associated with small transverse fluctuations. We believe that
the departure from extremality excites these massless modes, making
the 3-brane a dynamical object moving inside the 5-branes. 
Thus, the near-extremal
Bekenstein-Hawking entropy should coincide with that of a gas
of 3-branes of tension $T_{3\ eff} \sim   L_6 L_7 T_5$ moving in $5+1$ dimensions.

The near-extremal entropy of
\sol\  with $n_2=n_3=0$  ($H_2=H_3=1$) 
is given by \pop,\tre\ 
\eqn\yyu{
S = {8\sqrt\pi\kappa\over 3\sqrt {3 \prod_{i=1}^7 L_i}}
\sqrt{Q_1}\  E^{{3\ov 2}}\ . }
Using  \charges\  we rewrite this as 
\eqn\newb{
S = {4\sqrt{\pi n_1}\over 3 T_{3\ eff} \sqrt {3V}  }\ E^{{3\ov 2}}\ , }
where 
\eqn\deff{
T_{3\ eff} =  L_6 L_7 T_5  
\ , \ \ \ \ \ \ \  V =L_1 L_2 L_3 L_4 L_5  \ . }
{}For this formula to be applicable, the entropy has to be
macroscopic, i.e. $S\gg 1$. This puts a lower bound on $E$.
If we assume that $L_6$ and $L_7$ are of order $l_P\sim \k^{2/9}$, the
11-dimensional Planck scale, then we find
\eqn\lowerb{E \gg \left (V\over n_1 l_P^8\right )^{1/3} \ . 
}
At the same time, there is also an upper bound on $E$ coming from the
fact that the 5-brane is near extremality, $E\ll M_0$, which
implies
\eqn\upperb{
E\ll {n_1 V\over l_P^6} \ . 
}
When either $n$ or $V/l_P^5$ is large, there is a large range
of values of $E$ compatible with both \lowerb\ and \upperb.

We note that $S(E)$ in \newb\ grows faster than for
a gas of strings. {}For a $d+1$-dimensional free field theory,
$S(E)$ grows as $E^{d\over d+1}$ which is slower than for a 
gas of strings. This leads us to believe that the behavior
\newb\ can be explained only by dynamical $p$-branes with
$p>1$. Unfortunately, there is no known exact quantum theory for
such objects. The best one can do is to
study thermodynamics of $p$-branes 
in the semiclassical approximation 
(see, for example, \refs{\ortin,\byts}),
where one finds that, for sufficiently high energy,
$S\sim E^{2p\over p+1}$.

Thus, the entropy \newb\  is consistent with  that of a gas of 3-branes.
Since the growth of $S$ is faster than linear, the 
microcanonical ensemble is dominated by
a single large 3-brane.\foot{Clearly,
this maximizes the entropy for a
given total energy: dividing one $p$-brane of energy $E$ into
two $p$-branes of energy $\ha E$ would reduce the entropy by a factor
$2^{2\over p+1}$.} The thermodynamics here is highly unusual since the
temperature falls off with increasing energy as $1/\sqrt E$.
Note, however, that to compare with the Bekenstein-Hawking
entropy, \newb, it is sufficient to work in the microcanonical
ensemble, which is well-defined even for $p$-branes with $p>1$.  

{}For a semiclassical 3-brane, the world volume is described by
a $3+1$-dimensional theory containing some number, $m$, of massless
bosons and the same number of massless fermions. If the internal volume
of the 3-brane is $v$, and the world volume energy is
$e$, standard statistical mechanics gives
\eqn\stan{
S ={\textstyle{ {2\ov 3}} } \sqrt \pi (m v)^{1/4} e^{3/4} \ . }
In the semiclassical (gaussian) approximation 
the connection between the target space energy, $E$, and
the world volume energy is similar to that for strings,
\eqn\stri{
E^2 = x e \ .   }
{}For strings the exact quantum theory is available, and one finds
that $x\sim T_1 v$.
{}For 3-branes \stri\ holds in the semiclassical approximation (with $x \sim 
v^{1/3}T_3^{1/2} $), 
but we do not yet know how to determine $x$ in the exact theory.

We could turn the logic around and, assuming that the
Bekenstein-Hawking entropy is explained by the fluctuating 3-branes,
try to learn something about their quantum theory.
{}First of all, \newb\ implies that the relation \stri\ holds
in the full quantum theory, not just in the semiclassical
approximation. {}Furthermore, the matching of entropy requires that
\eqn\need{
x\sim v^{1/3} T^{4/3}_{3\ eff} V^{2/3} \ . }
The major difference from the case of strings is that this
relation involves the target space volume. 
This suggests that the 3-brane theory contains strong space-time
infrared effects. Perhaps such effects renormalize the 3-brane
tension from its bare value, $T_{3\ eff }$.

String theory ($p=1$) seems to be
the only case where such effects are not present, which
is intimately related
to the conformal invariance on the world sheet. Indeed, for $p=1$ neither
$x$ nor $S(E)$ depends on the target space volume, $V$. 
{}For the other solvable case, a gas
of particles ($p=0$) in $d+1$ dimensions,
$S\sim V^{1\over d+1} E^{d\over d+1}$. Here $V$ enters 
the entropy with a positive power,
and therefore it is natural that for 3-branes it enters
with a negative power. It is not yet clear to us why this power is
$-\ha$.

\newsec{Some Remarks on  `the 3-brane within a 5-brane' }

The appearance of a 3-brane  in $D=11$ at the intersection of two 5-branes
suggests a connection to the 3-brane of type IIB theory.
This connection can be made 
more explicit as follows.\foot{Note that there
 is also another  $D=11$ counterpart  of the
type IIB 3-brane. This is the  
$2_2$ configuration, i.e. the 2-brane with 
two  isometric transverse dimensions or, equivalently,  
 $2\bot 2$ with the second charge set equal to zero
(this configuration was already mentioned in section 2.3).}
Dimensional reduction of the $5\bot 5$ along one of the common
3-brane directions gives the $4\bot 4$ configuration (with a common 2-brane) 
in the type IIA theory. 
$T$-duality relates this to a `2-brane within 6-brane' configuration,
$6\Vert 2$.\foot{For a discussion of 
intersecting p-brane solutions  in $D=10$ and T-duality 
see \refs{\at,\beh}.}
$4\bot 4$ is also  $T$-dual to  
the $5\bot 3$ configuration of type IIB, 
i.e. to a D5-brane intersecting a D3-brane over a 2-brane. 
At the same time,  $T$-duality along one of the
common 2-brane directions 
transforms $4\bot 4$  into $(3\bot 3)_1$, i.e. $3\bot 3$   with one 
compact transverse direction. Both  $4\bot 4$
and $(3\bot 3)_1$ dimensionally reduce to the
$3\bot 3$ configuration of the $D=9$ theory.
Setting the second charge to zero gives $3_2$ in $D=9$.

In the reductions mentioned above, the intersection becomes
a 2-brane, while in the original M-theory configuration the
two 5-branes intersect over a 3-brane. We may ask whether it is 
possible to reduce to string theory in such a way that the intersection
remains a 3-brane. The answer is yes, but the reduced configuration
involves not only D-branes, but also NS-NS charged branes.
For example, we may dimensionally reduce $5\bot 5$ along  
one of the two  compact internal  
dimensions of the second 5-brane  which are 
orthogonal to the first one ($y_6$ or
$y_7$ in \sol).
This gives the 
$5\bot 4$ solution of type IIA theory, i.e. 
an NS-NS  5-brane intersected by a R-R 4-brane over a  
3-brane.  
 Since the 4-brane direction orthogonal to the 5-brane, $y_6$, is 
an isometry, we may apply  $T$-duality along this direction
to  obtain the 
$5_1\Vert 3 $ solution of type IIB theory\foot{
$T$-duality along one of the common directions
gives a D3-brane intersecting a NS-NS 5-brane over a 2-brane.
$T$-duality along one of the 5-brane directions that does not belong
to the D4-brane produces 
a D5-brane intersecting a NS-NS 5-brane over a 4-brane ($SL(2, R)$
transformation  interchanges the two 5-branes).},
i.e. a 3-brane lying within 
a 5-brane, with one of the transverse dimensions
being periodic.

This $5_1$ configuration is the Kaluza-Klein  5-brane  of type IIB theory
on $M^9 \times S^1$ \john\ (i.e. the
object magnetically charged under the Kaluza-Klein gauge field)
 which  is 
a singlet under the $SL(2, R)$
symmetry.
Its connection with the $5_2$ background in $D=11$
may be seen as follows.
We may start with \sol\  and set
$H_1=H,\  H_2=H_3=f=1$. The corresponding metric and 3-form field strength are
\eqn\yty{
ds^2_{11} = H^{2/3}\big [ H\inv  (-dt^2+ dy_n dy_n) + 
dy_6^2+ dy_7^2+ dx_i dx_i  \big ]  ,  \ \ 
 {\cal F}_4
 =3*dH \wedge dy_6\wedge dy_7  , }
where $n=1,...,5;\  i=1,2,3;\  r^2 = x_ix_i$.
Compactifying to $D=10$ along $y_7$ we get
the NS-NS 5-brane  $5_1$ 
of type IIA theory  in $M^9\times S^1$ described by 
the   $SO(3)$ symmetric  string sigma-model,
\eqn\opo{
L_{10} =  -\del t \bd t +  \del y_n \bd y_n   + 
 H(x)  (\del y_6 \bd y_6 + \del x_i \bd x_i )
+ B_i (x) (\del y_6 \bd x^i - \bd y_6 \del x^i)\ , }  
where $dB=-*dH$ and the dilaton is $\p=\ha \ln H$. $T$-duality along $y_6$ converts this 
`$H$-monopole' sigma-model into the
Kaluza-Klein monopole sigma-model 
with no dilaton and no antisymmetric 2-tensor and 
 the following non-diagonal metric,
\eqn\meet{
ds^2_{10} = -dt^2 + dy_n dy_n  + H\inv (x) (d\td y_6 + B_i dx^i)^2
+ H(x) dx_i dx_i\ . }
This is the metric 
describing  the  Kaluza-Klein 5-brane of type IIB theory.
  Note that the  direct lift of this metric into $D=11$ gives
a Kaluza-Klein 6-brane which  becomes the type IIA 6-brane upon 
reduction along $y_6$ \town.

Dimensional reduction of $5_1\Vert 3 $ along $y_6$ gives 
the $5\Vert 3$
configuration in 9 dimensions (upon further reduction to $D=4$
this in turn gives a black hole  \met\ in $D=4$ with 2 charges).
Thus, 
in the $D=9$ theory we find a D3-brane inside a  Kaluza-Klein 
5-brane.
If we  switch off the charge of the 3-brane, we 
return to the $5_1$ configuration in $D=10$, which reduces to
the Kaluza-Klein 5-brane in $D=9$.
We believe that the 3-brane interpretation
of the entropy of the latter can now  be rephrased  in terms of 
a fluctuating 3-brane of type IIB theory which lies within the 
Kaluza-Klein 5-brane. Unfortunately, 
since the Kaluza-Klein 5-brane is a purely NS-NS object,
this situation cannot be described
in D-brane terms.

Let us emphasize that the 5-branes in 11, 10 and 9 dimensions
have different scaling of the near-extremal entropy.
{}For the $D=11$ 5-brane
the scaling is that of a massless field theory;
for the $D=10$ 5-brane, which originates from the $5_1$ configuration
of M-theory, it is that of a string; for
the $D=9$ 5-brane, which originates from $5_2$, it is that of
a 3-brane. This variety of `non-critical' $p$-branes in $D=6$
may be related to type IIB compactifications on K3. At the point
in moduli space where a 2-cycle degenerates
a 3-brane wrapping around it gives rise to a 
tensionless string \Witten,
while a 5-brane gives rise to a tensionless 3-brane.
These objects  may be  related to the non-critical strings
and 3-branes that we are finding.


The low-energy fluctuations of the M-theory 5-brane are known
to be described by the $N=2$ tensor multiplet in $D=6$, whose
bosonic fields are $B_{mn}^-$ and 5 scalars. The scalars
have the interpretation of the transverse positions of the 5-brane in
$D=11$. If one is interested in the $5_1$ configuration in $D=11$,
which is characterized by non-critical strings, then one
of the scalars needs to be taken compact. This was indeed the case
for the quantization procedure of \dvv.
If we are interested
in the $5_2$ configuration, then two of the scalar fields need to be
made compact. Perhaps such a world-volume theory has a 
BPS saturated soliton solution
describing the embedded 3-brane. This would be similar in spirit to the
construction of \polc\ where the 3-brane soliton of $D=6$
supersymmetric abelian gauge  theory was found.


Our eventual goal is to derive an effective action for the
3-brane in $D=6$, similar to the $\kappa$-supersymmetric action in 
\polc, and below we speculate on the possible form of this action.
A clue may come from consideration of
an effective action for the
fluctuations of the $5\bot 5$ configuration in 11 dimensions
(with the second charge 
eventually set equal to zero). 
The first step towards constructing actions describing intersecting
M-brane configurations should be to identify 
the corresponding zero modes (as
was done for the basic 5-branes in \kap).
Since the extremal  $5\bot 5$  solution 
breaks  24 out of 32 supersymmetries 
of 11-dimensional theory we expect to 
find 24 (Goldstone)  fermionic zero modes and hence  12  bosonic zero modes. 
The `anisotropic 7-brane' form 
of the extremal $5\bot 5$ metric (\sol\ with $H_3=1, f=1$) implies
  the existence of  four normalizable  translational  zero modes, i.e. 
collective coordinates $X^i(\xi)$ ($i=1,...,4$). 
In addition, there should be  8 bosonic zero modes  $Z^p(\xi)$
($p=1,...,8$)
coming from  the antisymmetric 3-tensor (and related to breaking of the 
corresponding gauge symmetry). 
The absence of the full
 Lorentz symmetry of the 7-brane implies 
that the resulting    low-energy static-gauge 
action  will  have only the  reduced 
 $SO(1,3)\times SO(2) \times SO(2)$ 
global symmetry
\eqn\coll {
S_{5\bot 5} \sim \int d^8 \xi (  \eta^{\m\n} \del_\m X^i \del_\n X^i +
 c_1 \del_a X^i \del_a X^i + c_2  \del_s X^i \del_s X^i + ... ) \ . }
Here  $ \m,\n= 0,1,2,3; \ a=4,5; \ s=6,7$. 
The values of the  `anisotropy 
constants' $c_k$  depend on the values of  the two 5-brane charges $Q_k$ 
($c_1=c_2$ if $Q_1=Q_2$, $\ c_{1,2}=1$ if $Q_{2,1}=0$). 
The  $5_2$-brane action   should  
be a result of certain  (`double-dimensional') reduction  of \coll\ (with $
c_2=1$) along  $\xi^6,\xi^7$.

To   determine the  field content  of  the action for 
 the dynamical 3-brane moving inside the
5-brane  it is useful to  use supersymmetry considerations.
 The 5-brane theory has 16 conserved supercharges, and the presence
of a 3-brane breaks half of them. Thus, the
residual supersymmetry of the 3-brane
effective action is $N=2$ in $D=4$. This action must contain 8 Goldstone 
fermions, which naturally combine into two $D=4$ Majorana fermions.
By supersymmetry, the action should also contain 4 massless bosonic degrees 
of freedom.
We know that two of these bosons are the scalars 
describing the transverse fluctuations of the 3-brane in $D=6$.
One possible choice of 
the $N=2$ multiplet containing the required degrees of freedom 
is the vector multiplet. Then the low-energy effective action for
a single 3-brane is described by the $N=2$ supersymmetric $U(1)$ theory
(one may then speculate that multiple 3-branes are described by
the non-abelian $N=2$ vector multiplet).
Another possibility is to use the $N=2$ hypermultiplet. 
The two extra scalars
may then be interpreted as corresponding to the two compact coordinates
transverse to the 5-brane.\foot{
Note  also that the number of on-shell degrees of freedom, $4+4$,
  is  the same as in the $\k$-supersymmetric   action 
of a 3-brane moving in  8 space-time dimensions \ach.}   
It would be interesting
to study the 3-brane dynamics in $D=6$ in more detail.

\newsec{ Acknowledgements}
We  would like to thank  M. Cveti\v c, A. Hanany and  J. Maldacena
for useful discussions. 
I.R.K. was supported in part by DOE grant DE-{}FG02-91ER40671, the NSF
Presidential Young Investigator Award PHY-9157482, and the James S.{}
McDonnell {}Foundation grant No.{} 91-48.
A.A.T. would like to 
acknowledge  the support of PPARC,
ECC grant SC1$^*$-CT92-0789 and NATO grant CRG 940870.
\vfill\eject
\listrefs
\end